\documentclass[a4paper, 10pt]{article}
\usepackage{tabularx}
\usepackage{array}
\usepackage{graphics}
\usepackage{graphicx}
\usepackage{epsfig}
\usepackage{amsmath}
\usepackage{amsfonts}
\usepackage{amssymb}

\newcommand{\beq}{\begin{equation}}
\newcommand{\eeq}{\end{equation}}
\newcommand{\bea}{\begin{eqnarray}}
\newcommand{\eea}{\end{eqnarray}}

\begin{document}

%opening
\title{
\begin{flushright}
\begin{tabular}{l}
\small{\tt LPT-Orsay/08-44}\\
\small{\tt UHU-FT/0803}\\
\strut \\
\strut \\
\end{tabular}
\end{flushright}
\large\bf IR finiteness of the ghost dressing function from numerical resolution of  the ghost SD equation}
\author{Ph. Boucaud$^a$,  J-P. Leroy$^a$,  A. Le~Yaouanc$^a$,  J. Micheli$^a$, \\ O. P\`ene$^a$,  J. Rodr\'{\i}guez--Quintero$^b$}

\maketitle

{\par\centering \vskip 0.5 cm\par}
{\par\centering \textsl{ 
$^a$~Laboratoire de Physique Th\'eorique \footnote{Laboratoire associ\'e au CNRS,  UMR 8627}(B\^at.210),  Universit\'e de
Paris XI, \\ 
Centre d'Orsay,  91405 Orsay-Cedex,  France.} \\
\vskip 0.3cm\par }{\par\centering \textsl{ 
 $^b$ Dpto. de F\'{\i}sica Aplicada,  Fac. Ciencias Experimentales, \\Universidad de Huelva,  21071 Huelva,  Spain} 
\vskip 0.3cm\par }

\begin{abstract}

{\small We solve numerically the Schwinger-Dyson (SD hereafter) ghost equation in the Landau gauge for  a  given gluon propagator finite at $k=0$ ($\alpha_{gluon}=1$) and with the usual assumption of constancy of the  ghost-gluon vertex ; we show that there exist two possible types of ghost dressing function solutions,  as we have previously inferred from analytical considerations : one singular at zero momentum,  satisfying the familiar relation $ \alpha_{gluon}+2 \alpha_{ghost}=0$ between the infrared exponents of the gluon and ghost dressing functions(in short, respectively $\alpha_G$ and $\alpha_F$ \footnote{Let us recall that we denote the gluon by a G,  and then the  ghost by a  F,   for  "fant\^omes" }),  and having therefore $\alpha_{ghost}=-1/2$,  and another which is finite at the origin ($\alpha_{ghost}=0$) ,  which violates the relation. It is most important that the type of solution which is realized depends on the value of the coupling constant. There are regular ones for any coupling below some value,  while there is only one singular solution,  obtained only at a critical value of the coupling. For all momenta $k<1.5$~GeV where they can be trusted,  our lattice data exclude neatly the singular one,  and agree very well with  the regular solution we obtain at a coupling constant compatible with the bare lattice value.}

\end{abstract}

\section{Introduction}
Since the first attempts,  an impressive progress has been made in understanding the solutions to the Schwinger-Dyson equations for the QCD propagators and their behaviour at small momenta. In particular,  an important step has been accomplished by putting forward the essential contribution of internal ghost loops in the gluon propagator equation,  previously neglected ; it has been shown that it may completely change the previously expected behavior
of the gluon propagator from much more singular than the free one ( something like $1/k^4$- as was believed for a long time),  to being much  less singular\cite{alkofer-AnnPhys1998} \footnote{For this particular case,  see especially the section 3 of the quoted paper}. All the following considerations are assuming the choice of the Landau (i.e. Lorentz) gauge. 
\vskip 0.2cm
-{\bf The consensus before 2005} 

For some years,  a consensus seemed to be obtained around a statement of 1) an infrared fixed point of the gluon-ghost coupling 2) a singular ghost dressing function (see below for more explanation). This consensus was very strong and unopposed,  since several other approaches were apparently converging to the same conclusions. Such authoritative people as S. Brodsky have also appealed to it to support their considerations about AdS/CFT ( see \cite{brodsky-ph-0703109}\footnote{See especially the section 3 of the paper by Brodsky giving references to certain lattice data and to non perturbative statements like solutions of SD equations}). Another part of the consensus, 
deduced from a solution to coupled SD equations,  was 
the statement that $\alpha_G>1$,  i.e. the gluon propagator should necessarily vanish;  yet this was contested by a thorough calculation of Bloch \cite{bloch-ph-0303125}.

To be more specific about this consensus, a usual assumption for the infrared behaviour of gluon  and ghost dressing functions is that they should be power behaved,  i.e. for the gluon $G(k)$ and the ghost $F(k)$  \textit{dressing functions}
 respectively :
\begin{eqnarray}
G(k) \sim (k^2)^{\alpha_G} \\
F(k) \sim (k^2)^{\alpha_F}
\end{eqnarray} 
In a number of studies \cite{alkofer-AnnPhys1998, zwanziger-th-0109224},   it has been stated that for a suitably simple assumption concerning the ghost-gluon and gluon vertices,  the Dyson-Schwinger (SD hereafter) coupled equations  for  $G(k) $ and $F(k) $ imply 
\begin{equation}
\alpha_G+2 \alpha_F=0 \label{IRstable}
\end{equation} 
This statement is the starting point for the popular claim of an infrared fixed point for the QCD renormalised coupling
constant. In fact,  admitting the validity of eq.  (\ref{IRstable}),  the IR fixed point  would be present in  the coupling
constant defined by the ghost-gluon 3 points Green function in a MOM  scheme. Let us recall the renormalisation conventions, 
with bare quantities denoted by a "B" subindex. In general :
\begin{eqnarray}
G_B(k^2)=Z_3~G_R(k^2), ~
F_B(k^2)=\widetilde Z_3~F_R(k^2), ~
g_B=Z_g~g_R, \nonumber \\
\Gamma_R=\widetilde z_1~\Gamma_B, ~
Z_g=\widetilde z_1~(Z_3^{1/2}~\widetilde Z_3)^{-1}
\end{eqnarray}
In the MOM schemes :
\begin{eqnarray}
Z_3=G_B(\mu^2)\\
\widetilde Z_3=F_B(\mu^2)\\
G_R(k^2,\mu)=G_B(k^2)/G_B(\mu^2) \\
F_R(k^2,\mu)=F_B(k^2)/F_B(\mu^2)
\end{eqnarray} 
 while many possibilities are opened  for the renormalisation condition of the vertex. We need not specify it for reasons explained,  below eq. (\ref{g2eff}).
Then:
\begin{equation}
g_R(\mu)=g_B G_B(\mu^2)^{1/2} F_B(\mu^2) /\widetilde z_1(\mu) 
\end{equation}
This implies that the product $g_R(\mu)\widetilde z_1 (\mu) (G_R(k^2, \mu))^{1/2} F_R(k^2, \mu) $
is independent of $\mu$.

Now,  $g_R(\mu)$ would tend to a finite limit for small $\mu$ if  eq.  \ref{IRstable} would hold,  under the additional assumption that $\widetilde z_1(\mu)$ is finite for  $\mu \to 0$ \footnote{Note that the UV finiteness of $\widetilde z_1(\mu)$ does not imply that is $\mu$ independent,  even in perturbation,  and in particular for the symmetric MOM scheme,  see eq. 13 in our ref.  \cite{nous-ph-0507104v4},  extracted from the results of Chetyrkin and Retey \cite{chetyrkin}; however, we can suppose that the non perturbative IR behavior is not too singular.}.
\vskip 0.2cm

-{\bf The input of lattice data} 

Recently,  lattice data have also entered the game and have contributed much to the discussion,  by showing  features quite contrary to this consensus. Our motivation here is to try to clarify the situation within the SD approach by exhibiting new numerical solutions restoring the agreement between the lattice data,  the numerical study in the continuum,   and the analytical considerations \footnote{An attempt to describe the lattice data within SD coupled equations is made in \cite{aguilar-ph-0408254}. For a recent attempt to accommodate the lattice data (with a finite non zero ghost dressing function) within the Gribov-Zwanziger approach, see \cite{dudal-th-0711.4496}}.

We can test the relation (\ref{IRstable}) on the lattice,  by computing $G(k) F(k)^2$,  which according to the above relation (\ref{IRstable}) would be expected to tend to a finite value at small $p$. 
In fact,  this is clearly contradicted by the lattice data,  see Fig. 1 in our paper \cite{nous-ph-0507104v4},  and,  from Sternbeck et al. Fig. 4 in \cite{sternbeck-lat-0506007},  Fig.3 in \cite{sternbeck-lat-0511053},  which show that the product {\bf decreases rapidly} at small $p$,  possibly to zero. In addition, as we will see from sections \ref{analytical} and \ref{numerical}, for solutions satisfying the relation (\ref{IRstable},the product $N_c g_R(\mu)^2 \widetilde z_1 (\mu) G_R(k^2,\mu) F_R(k^2,\mu)^2 $  must tend, when $k \to 0$, to a finite value which is much larger than the value observed at the smallest accessible momenta (it should be $10~\pi^2$ in the case $\alpha_G=1$). 

In addition,  in the above studies,  which  try to solve the \textbf{coupled} SD equations,  the gluon propagator is also predicted,  and it is found that $\alpha_G$ is positive,  which is anyway also suggested by lattice QCD. If so,  and if the relation (\ref{IRstable}) would hold, then it would imply finally that $\alpha_F<0$,  i.e. the ghost dressing function should be singular. One could consider this  result on the ghost dressing function to be theoretically appealing,  because it meets another familiar statement,  the Kugo-Ojima criterion for confinement (according to which the ghost dressing function should be singular).

But,  from the admitted values of $\alpha_G \gtrsim 1$,  $\alpha_F \lesssim -1/2$,  $F(k^2)$ should present a power behavior close to $1/k$,  or more singular. This stronger result is excluded by the lattice data,  which allow at most a \textbf{very weak} singularity ; indeed according to our first analysis \cite{nous-ph-0507104v4} and to the study of Sternbeck et al. \cite{sternbeck-lat-0610053}(see also their recent large volume study \cite{bogolubsky-lat-0710.1968}),  the power seems at most $\alpha_F=-0.2$ down to momenta around $0.3$~GeV \footnote{Note that these authors plot $q^2$ along the $x$ axis in the Figures}; in fact,  we have obtained better fits of our own $SU(2)$ and $SU(3)$ data  with logs rather than with powers \cite{nous-ph-0604056}; finally,  if we abandon any prejudice,  it appears that it is compatible with a finite value as well. Certainly,  something close to $\alpha_F=-0.5$ is not possible,  unless there is a sudden change of behavior very near $k=0$. This conclusion is reinforced by the recent results of Cucchieri and Mendes at very large volumes \cite{cucchieri-lat-0710.0412}.
\vskip 0.2cm

-{\bf Analytical setting} 

In view of this situation,  in our paper \cite{nous-ph-0507104v4},  we started a new discussion on the implications of the ghost SD equation for the IR behavior of the ghost propagator.

We consider this equation in its {\bf subtracted,  UV convergent} form:
\bea\label{SD}
\frac{1}{F_B(k^2)}-\frac{1}{F_B({k^{\prime}}^2)}=-N_c g_{B}^2 \int\frac{d^4 q}{(2\pi)^4}
\left( 1-\frac{(k.q)^2}{k^2 q^2} \right) \nonumber \\
\left[\frac{G_B((q-k)^2)H_{1B}(q, k)}{((q-k)^2)^2}-\frac{G_B((q-k^\prime)^2)H_{1B}(q, k^\prime)}{((q-k^\prime)^2)^2}\right] F_B(q^2)
\eea
$k^\prime$ is an arbitrary subtraction point,  taken for simplicity parallel to $k$,  $k^\prime =k \sqrt{{k^\prime}^2/k^2}$.
$H_1$  is one of the invariants in the Lorentz decomposition of the ghost-gluon vertex:
\bea
\widetilde{\Gamma}^{abc}_{B\mu}(-q, k;q-k)&=&i g_B f^{abc} q_\nu\widetilde{\Gamma}_{B\nu\mu}(-q, k;q-k) \nonumber \\
&=&i g_B f^{abc} (q_\mu H_{1B}(q, k)+(q-k)_\mu H_{2B}(q, k))
\eea
where $-q$, $k$ and $q-k$ are respectively the entering momenta of outgoing ghost, ingoing one and gluon, .

In all our present  considerations  this equation is considered for  given gluon propagator and vertex,  and the ghost dressing function appears then as the solution to the equation.  This is what we call the SD ghost equation. We do not try to solve any other SD equation.
The advantage
of concentrating on this equation is that it is much simpler than the gluon one or any other,  to the point that analytical statements can be formulated for the ghost,   with given gluon propagator and vertex. On the other hand,   various assumptions on the IR behavior of the gluon propagator and the vertex may be used,  in particular those advocated in the above references. Both these inputs and the output ghost solution can be  tested through the lattice data. 

The renormalised form of this equation is :

\bea\label{SDR1}
        \frac{1}{F_R(k^2)}-\frac{1}{F_R({k^\prime}^2)}=-N_c g_{R}^2 \widetilde{z_1} \int\frac{d^4q}{(2\pi)^4}
\left(1- \frac{(k.q)^2}{k^2 q^2} \right)  \nonumber \\
\left[ \frac{G_R((q-k)^2)H_{1R}(q, k)}{((q-k)^2)^2}-\frac{G_R((q-k^\prime)^2)H_{1R}(q, k^\prime)}{((q-k^\prime)^2)^2}\right] F_R(q^2)
\eea
where $\widetilde{z_1}$ is the renormalisation constant of the ghost-gluon vertex. We know that,  in Landau gauge,  $H_{1B}(q, 0)+H_{2B}(q, 0)=1$ which implies that $\widetilde{z_1}$ is
finite for any momentum configuration \footnote{Let us recall that what has been really demonstrated in the paper of Taylor \cite{taylor} is the equation we have just written,  i.e. for a vertex with zero ingoing ghost momentum. Then $\widetilde{z_1}=1$ for this particular MOM renormalisation ; in general it will remain finite but different from $1$. The detailed explanations on the Taylor paper are given in our article \cite{nous-ph-0507104v4}}. 
Let us  remark that this implies that  the subtracted SD equation is convergent. Indeed,  $\widetilde{z_1}$ and the l.h.s. of eq.(\ref{SDR1}) being finite the integral in the r.h.s. must be convergent. This was not obvious in the bare version.

In the following,  we set ${k^{\prime}}^2=\mu^2$ to get the one variable renormalised integral equation.

One can wonder whether the solutions of this subtracted SD equation are also solutions of the unsubtracted one :
\begin{eqnarray} \label{Runsubtracted2}
\frac{1}{F_R(k^2)}=\widetilde Z_3-N_c g_{R}^2 \widetilde z_1 \int\frac{d^4 q}{(2\pi)^4} 
\left(1- \frac{(k.q)^2}{k^2 q^2} \right)  \nonumber \\
\left[ \frac{G_R((q-k)^2)H_{1R}(q, k)}{((q-k)^2)^2}\right] F_R(q^2)
\end{eqnarray}
This is seen to hold simply by making  :
\begin{align} \label{Z3}
\widetilde  Z_3&=1+
N_c g_{R}^2 \widetilde{z_1}   \nonumber \\ 
\int&\frac{d^4q}{(2\pi)^4}
\left(1- \frac{(k.q)^2}{k^2 q^2} \right)
 \left[\frac{G_R((q-k)^2)}{((q-k)^2)^2}\right] H_{1R}(q, k) \left. F_R(q^2) 
\right\arrowvert_{k^2=\mu^2}
\end{align}
Of course,  one has now to regularise the integral in some way,  and this introduces a finite arbitrariness in $\widetilde  Z_3$.
The divergence of $\widetilde  Z_3$,  coming from the integration,  will be of course cancelled, as it was in the subtracted form (\ref{SDR1}), by the divergence of the remaining part of the integral.
\vskip 0.2cm

-{\bf Introducing the regular solutions for the ghost dressing function} 

 We concluded in ref. (\cite{nous-ph-0507104v4}) that,  \textbf{in general},  under the usual IR regularity assumption for the ghost-gluon vertex,  the  SD ghost equation implies  the relation (\ref{IRstable}) by itself,  without recoursing to the gluon equation. However,  there were exceptions (see below),  but we have first discarded them. Therefore,  since the relation is definitely seen to be violated on the lattice,  while the SD equation is automatically satisfied,  we first suggested in the same paper,  as a way out of this puzzle,  that the vertex invariant $H_1$ could be IR singular, 
instead of being constant. 

Then,  it soon appeared,  in view of the lattice data,  in particular thanks to Sternbeck et al. - \cite{sternbeck-lat-0511053} as well as to the previous work of Cucchieri et al.  \cite{cucchieri-lat-0408034},  that this possibility is very unprobable : indeed  they measure $H_{1B}(q, q)$ (gluon at zero momentum,  and contraction with $q_{\mu}$),  and they find it roughly constant and close to 1. Therefore,  our attention has been drawn to the cases predicted  in our analytical discussion of the SD ghost equation \cite{nous-ph-0507104v4},  where the relation \ref{IRstable} can be  violated in spite of having a regular ghost-gluon vertex. These are the cases where $\alpha_G \simeq 1$ and $\alpha_F=0$,  i.e. where {\bf  the ghost dressing function is regular \footnote{For some qualification of the term "regular" used in the present context, see the Introduction, below eq. (\ref{ghostlogs})} at origin}. As we have said,  we did not pay attention to them in the beginning. But we have become aware that this possibility is attractive because :

1) on the lattice,  $\alpha_G$ seems not far from $1$,\cite{bonnet-lat-0101013, nous-lat-0602006},  i.e. the gluon propagator is not far from being finite (see also the very recent very large volumes studies of the above references \cite{bogolubsky-lat-0710.1968},\cite{cucchieri-lat-0710.0412}). Thus it automatically leads to a rapidly decreasing $G(k^2) F(k^2)^2$, ${\cal O}(k^2)$, behaviour when $k$ approaches $0$ in agreement with what is observed. 

2) last but not least, on the lattice, the effective $\alpha_F$ is compatible with $0$, as we have seen above.

Therefore,  the appealing possibility $\alpha_F=0$,  has been adopted in our subsequent paper  \cite{nous-ph-0604056}. 

On the other hand,  no statement  can be made from these analytical considerations on the ghost SD equation as to which solution for the ghost propagator
should be effectively preferred  in real QCD. 
{\bf A complementary theoretical input comes from the Slavnov-Taylor (ST) identity for the three-gluon vertex}. From this identity,  we have demonstrated in \cite{nous-ph-0701114} that the ghost dressing function should be IR finite. 

The aim of the present paper is to reconsider  this question by a {\bf numerical} study of the ghost SD equation with input from lattice data for the gluon propagator,  and the simple and widely admitted constancy assumption for the vertex. The conclusion is striking : it is found that  the  IR finite solutions violating the relation (\ref{IRstable}) indeed exist. Which  type of solution is realized,  either singular or regular at $k \to 0$,  depends on the actual value of the QCD coupling constant. {\bf One} singular solution and {\bf only one} is obtained,   for only {\bf one}  value of $g^2$ which we call "critical",  and it can be completely exhibited ;  according to our calculation,  {\bf  it cannot  be the one of real QCD,  because it disagrees  grossly with the lattice results over a large range of momenta}. Therefore, in agreement with our ST statement,  the actual ghost dressing function must be regular,  i.e. IR finite,  and indeed {\bf we find solutions regular at $k \to 0$   describing very  well the lattice  ghost  data},  with values of the coupling constant close to the one estimated from the actual bare coupling constant of the lattice. In summary,  the combination of  the numerical resolution and of the ghost lattice data or the ST theoretical input allows to discard the singular  solution.
\vskip 0.2cm

-{\bf Warnings} 

A caveat must be made now. To make the discussion simple and to keep it close to the commonly accepted conceptual framework,  we have adopted above the usual assumption  of  a \textbf{pure power} IR behaviour for dressing functions. But this is by no means a  necessary assumption;  especially for a massless theory,  it would not be unexpected to have a  behavior with log factors accompanying integer powers of $k^2$ . For instance,  according to our Slavnov-Taylor discussion \cite{nous-ph-0507104v4, nous-ph-0701114, nous-ph-0702092}, 
the gluon propagator must be infinite at the origin, under some regularity assumption for the three-gluon vertex. Then a way to reconcile this statement with the observation of an apparently IR finite 
propagator from the lattice is to assume that this infinity is logarithmic,  which would make it very difficult to detect on the lattice \footnote{Of course,  the gluon propagator is always finite at $k=0$ in a finite volume. What we mean here is that the gluon propagator at $k=0$ seems to be relatively constant for increasing volume. And,  one cannot exclude a very slow variation,  such as expected from a logarithm.}. The behavior of $G(k)$ would be $G(k) \simeq k^2 (\log(k^2))^{\nu}$ ($\nu >0$). We present the results disregarding the logs. We have also checked in our numerical calculation that {\bf including such a log in the gluon propagator does not
change appreciably the ghost propagator} deduced from the SD equation.

In addition,  in our analytical discussion \cite{nous-ph-0701114},  we have shown that if $\alpha_G=1$ and $\alpha_F=0$,   one must have in the ghost dressing function logarithms of the type 
\bea \label{ghostlogs}
F(k)=a+b~k^2 \log(k^2).
\eea 
The effect of such logarithms is very weak, 
so that for the present purpose we count this as "regular",  and anyway,  it {\it is} IR finite. Nevertheless,  we are able to display the effect of this logarithm in the numerical calculation.

As we have said,  we have checked that including a log in the gluon propagator does not
change appreciably the ghost propagator deduced from the SD equation. 
This is in agreement with an analytical discussion,    which shows that only the power of the log in eq.(\ref{ghostlogs} ) is changed. 

\vskip 0.2cm

-{\bf Assumptions and inputs of the calculation} 

To summarise,  our starting assumptions for the numerical calculation below are finally the following :

1) we take the  ghost gluon vertex invariant $H_1(k.q)$ implied in the  SD ghost equation (see below) as momentum independent. This is  a rather usual assumption made in SD studies,  and as we said above,  it is in rough compatibility with
present lattice data. In fact,  it would be sufficient to assume simply a regular vertex to get the same qualitative conclusions,  but for a numerical study we have to make a definite choice.  Anyway,   let us emphasize that our first goal is not to make a realistic quantitative prediction,  but  to demonstrate that, 
even with this type of regularity assumption,  one can obtain solutions not  obeying  $\alpha_G+2 \alpha_F=0$,  contrarily to the common belief. For this purpose,  it is not required to bother about what would be the most realistic assumption for the vertex.

2) for the gluon propagator,  in the small momentum region,  we use an interpolation of the gluon propagator given by the lattice,   with $\alpha_G = 1$. 

We are aware that several SD studies,  for example \cite{alkofer-AnnPhys1998},   exclude  this latter possibility when considering
the coupled equations,  since  $\alpha_G$ and $\alpha_F$ are then determined separately,  and they found then that $\kappa=-\alpha_F=1/2 \alpha_G >1/2$,  therefore  $\alpha_G >1$. At this point,  we recall that  the paper of Bloch \cite{bloch-ph-0303125} concerning the coupled equations  finds solutions with $\alpha_G =1$ and $ \alpha_G+2 \alpha_F=0$ ($\kappa=1/2$),  thanks to a more refined treatment of the gluon SD equation. He can then reproduce rather well the gluon lattice propagator. 
Then the question is compelling : knowing that the gluon lattice data are well reproduced by him,  and that the lattice data satisfy the ghost SD equation he is solving,  how can it be that the lattice ghost dressing function differ from his prediction $1/k$ ? We show that this is not due to lattice artefacts,  but to the neglected possibility that there are different type of solutions to the ghost equation for the {\bf same} gluon propagator,  depending on the value of the coupling constant.

\section{ Analytical considerations on the behaviour at small $k$}
\label{analytical}
As a preliminary to the numerical study,  let us recall or establish
{\bf analytical} relations,  which can be used as tests of the soundness and accuracy of our numerical calculation. We extract them from a more complete discussion,  which  will be given in a forthcoming paper \cite{nous-forthcoming}. 

In this section and hereafter,  since we adopt the constancy assumption for the ghost-gluon vertex,  it appears immediately that the coupling constant only appears in the combination  :
\begin{equation}
\widetilde {g}^2 \equiv N_c g_{R}^2 \widetilde{z_1} H_{1R} = N_c g_B^2 Z_3 \widetilde Z_3^2/\widetilde{z_1} H_{1R}= N_c g_B^2 Z_3 \widetilde Z_3^2 H_{1B} \label{g2eff}
\end{equation}
where $ H_{1R}$ or $H_{1B}$ are constants. We use this auxiliary notation througout the rest of the article. From the last equality,  it is obvious that $\widetilde {g}^2$  is independent of the way one renormalises the vertex,  since only the bare vertex takes place.

We proceed as in our paper \cite{nous-ph-0507104v4} : 
- we separate the integral into one UV part and an IR part,  respectively for $q>q_0$ and $q<q_0$.

-for the infrared contribution (see eq.(14) and (17) of \cite{bloch-ph-0303125} with $\alpha_{\Gamma}=0$ and
 $H_1$ and $h$ constant),  defining $A$ and $B$
by $F_R(k)\simeq A (k^2)^{\alpha_F}$ and $G_R(k)\simeq B (k^2)^{\alpha_G}$ when $k\rightarrow 0$,  i.e. for $k<q_0$.
 
-we write eq.(\ref{SDR1}) replacing $k$ by $\lambda k$,  taking $k^\prime=\lambda
\kappa k$ and performing the change of variable $q\rightarrow \lambda q$.
We then consider the IR limit $\lambda \to 0$. 

1) {\bf Singular solution}. Let us establish a relation for $\widetilde {g}^2_c$,  the value of $\widetilde {g}^2$ corrresponding to the singular solution,  which is found to be unique.
It is inspired by Bloch \cite{bloch-ph-0303125}.
 
0ne can write,  at leading order in $\lambda$ :

\bea
(\lambda^2 k^2)^{-\alpha_F} (1-\kappa^{-2 \alpha_F})\simeq -\widetilde {g}^2_c (\lambda^2)^{\alpha_F+\alpha_G} A^2 B \int^{q<  {q_0}/{\lambda}}\frac{d^4 q}{(2\pi)^4}(q^2)^{\alpha_F}\left(1-\frac{(k.q)^2}{k^2 q^2}\right)  \nonumber \\
\times\left[((q-k)^2)^{\alpha_G-2}-((q-\kappa k)^2)^{\alpha_G-2}\right]
\eea

This integral being ${\cal O}(\lambda^{2(\alpha_F+\alpha_G)})$ dominates over the UV part which is ${\cal O}(\lambda^2)$,  by a negative power of $\lambda$, for $\alpha_G=1$ and $\alpha_F<0$. We can then neglect the UV part. 
This gives the relation $2\alpha_F + \alpha_G = 0$, whence $\alpha_F=-1/2$, $F(k^2) \simeq 1/k$. Moreover the integral of the r.h.s. can be analytically performed. Defining the function:
\beq
f(a, b)=\frac{1}{16\pi^2}\frac{\Gamma(2+a)\Gamma(2+b)\Gamma(-a-b-2)}{\Gamma(-a)\Gamma(-b)\Gamma(4+a+b)} 
\eeq
its value is equal to $(1-\kappa^{-2\alpha_F})(k^2)^{-\alpha_F}\Phi(\alpha_G)$ where:
\bea
&\Phi(\alpha_G)=& \nonumber \\
&-\frac{1}{2}\Big(f(-\frac{\alpha_G}{2}, \alpha_G-2)+f(-\frac{\alpha_G}{2}, \alpha_G-1)+f(-\frac{\alpha_G}{2}-1, \alpha_G-1)\Big)+ \nonumber \\
&\frac{1}{4}\Big(f(-\frac{\alpha_G}{2}-1, \alpha_G-2)+f(-\frac{\alpha_G}{2}-1, \alpha_G)+f(-\frac{\alpha_G}{2}+1, \alpha_G-2)\Big) 
\eea
leading to $\widetilde {g}^2_c A^2 B=\frac{1}{\Phi(\alpha_G)}$. In our case,  $\alpha_G=1$ and $\alpha_F=-\frac{1}{2}$,  $\Phi(1)=\frac{1}{10\pi^2}$ and the relation becomes:

\beq\label{g2c}
\widetilde {g}^2_c A^2 G_R^{(2)}(0)=10\pi^2
\eeq
where $G_R^{(2)}$ is the gluon propagator.

This is only a relation between  $\widetilde {g}^2_c$ and $A$ and it doesn't allow us to know a priori the value of $\widetilde {g}^2_c$ before any numerical computation,  unless a very small renormalization point $\mu$ is choosen. In this case we have:
$A=(\mu^2)^{-\alpha_F}$ and $B=(\mu^2)^{-\alpha_G}$ so that $\widetilde {g}^2_c=\frac{1}{\Phi(\alpha_G)}$ ($\widetilde {g}^2_c=10\pi^2$ when $\alpha_G=1$).

2) {\bf Regular solutions}. If,  on the other hand,  $\alpha_F=0$,  then the l.h.s. is trivially zero at leading order in $\lambda$,  and one has to go a step further in the expansion to get a non trivial result. Noting that $A=F_R(0)$ and $B=G_R^{(2)}(0)$,   where $G^{(2)}$ is the gluon {\bf propagator},  finite at the origin,  the IR part of the integral has the form :
\bea
-\widetilde {g}^2 \lambda^2 F_R(0)^2 G_R^{(2)}(0) \int^{q<  {q_{\mbox {\tiny  0}}}/{\lambda}}\frac{d^4 q}{(2\pi)^4} \left(1-\frac{(k.q)^2}{k^2 q^2}\right) 
\times\left[\frac {1} {(q-k)^2}-\frac {1}{(q-\kappa k)^2}\right]= 
\nonumber \\
-\widetilde {g}^2 \frac {1} {64 \pi^2}\lambda^2 k^2(1-\kappa^2) \log(q_0/(\lambda~k)) F_R(0)^2 G_R^{(2)}(0) +...
\eea
where the dots denote subleading ${\cal O}(k^2)$ terms.
It still dominates over the UV part,  although this time it is only by a logarithm. We then write consistently $F_R(k^2)=a+b k^2 \log(1/k^2)+{\cal O}(k^2)$  
 in the l.h.s,  and we find:
\beq\label{regular}
F_R(k^2)=F_R(0)\left (1-\widetilde {g}^2\frac {1} {64 \pi^2} F_R(0)^2 G_R^{(2)}(0) k^2 \log(M^2/k^2) \right)
\eeq
$M^2$ being some scale which we cannot derive from this IR expansion.

\section{Numerical solution of the Schwinger-Dyson equation for the ghost}
\label{numerical}

In this section we want to see if the two types of solutions ($\alpha_F=0$ and $2\alpha_F+\alpha_G=0$) suggested by our analytical discussion in \cite{nous-ph-0507104v4, nous-ph-0604056} actually exist for the same gluon propagator. We answer positively by solving numerically the ghost SD equation for  given gluon propagator and vertex.
In the following we shall use the subtracted form of the Schwinger-Dyson equation for the ghost in the Landau (i.e. Lorentz)  gauge. This equation has been written above,  eq. (\ref{SDR1}).

We start from an IR finite  gluon propagator ($\alpha_G=1$) extracted from our lattice data in  pure Yang-Mills theory,  with Wilson gauge action,   $\beta = 5.8$ and a lattice volume equal to $32^4$),  for momenta lower than $1.5$~GeV ; this choice is justified to have moderate UV artefacts. We  extend it  to  larger momenta using a one loop asymptotic expansion (with $\Lambda_{MOM}=1$~GeV corresponding to the standard $\Lambda_{\overline MS}=.240$~GeV of lattice quenched QCD).
On the other hand,  we take $H_1(q, k)$ to be constant with respect to both momenta \footnote{This cannot be an exact statement,  as already shown in perturbation by the calculations of ref. \cite{davydychev, chetyrkin} : although finite,   the vertex invariants do depend on the momenta through the running $\alpha_s$ }.   As we said above,  this is suggested by the lattice data for $q=k$ (i.e. for zero gluon momentum),  but we extend it to all values of $q$ and $k$. The authors of ref.\cite{sternbeck-lat-0511053} find a bare vertex very close to $1$ in this zero momentum gluon configuration for a large range of $\sqrt{q^2}$. 

We work in the $MOM$ scheme, and set ${k^\prime}^2$ appearing in eq.(\ref{SDR1}) as the squared renormalisation scale $\mu^2$ ($\mu$ has been chosen at an optimum $1.5$~GeV,  not too high to allow the lattice data to be safe,  and not too small to display the differences between solutions at small momenta).  The equation we have to solve is :
\bea\label{SDRcst2}
\frac{1}{F_R(k^2)}=1-\widetilde {g}^2 \int\frac{d^4q}{(2\pi)^4} 
\left(1-\frac{(k.q)^2}{k^2 q^2} \right)  \nonumber \\
\left[ \left.\frac{G_R((q-k)^2)}{((q-k)^2)^2}-\frac{G_R((q-k^\prime)^2)}{((q-k^\prime)^2)^2}\right]  F_R(q^2) \right\arrowvert_{{k^\prime}^2=\mu^2}
\eea
\vskip 0.3 cm
Note that  this equation implies that $\widetilde {g}^2$  depends only on the renormalisation point chosen for the propagators ; it is independent of the particular way used to define the renormalisation of the vertex,  in agreement with eq.(\ref{g2eff}). 
Eq. \ref{SDRcst2} can be still  transformed to a new form which makes the numerical calculation and the presentation of the various solutions easier ; for this,  we subtract the equation at $k=0$,  to let the value of $F_R(k)$ at the origin appear and to eliminate the reference to the particular normalisation point $\mu$,  and we redefine also the unknown function to be calculated as $\widetilde {F}(k)=\widetilde {g}F_R(k)$. Then the reference to the value of $\widetilde {g}$ also disappears; we end with :
\bea\label{SDRcst3}
\frac{1}{\widetilde {F}(k^2)}=\frac{1}{\widetilde {F}(0)}- \int\frac{d^4q}{(2\pi)^4}
\left(1- \frac{(k.q)^2}{k^2 q^2}\right)  \nonumber \\
\left[ \frac{G_R((q-k)^2)}{((q-k)^2)^2}-\frac{G_R((q)^2)}{((q)^2)^2}\right]  \widetilde {F}(q^2)
\eea
\vskip 0.3 cm

We solve this equation (\ref{SDRcst3}) for $\widetilde {F}(k^2)$, for a set of values of  $\widetilde {F}(0)$. It is easy to see that from this solution we can reconstruct the desired solution of eq. (\ref{SDRcst2}) for any renormalisation point and any value of $\widetilde {g}$. Indeed,  we 
have $\widetilde {g}(\mu)=\widetilde {F}(\mu^2)$,  so that for given  $\mu$ and $\widetilde {g}$,  we have just to identify the value of
 $\widetilde {F}(0)$ such that $\widetilde {F}(\mu^2)= \widetilde {g}$. Then,  we reconstruct  ${F_R}(k^2)$ through ${F_R}(k^2)= \widetilde {F}(k^2)/ \widetilde {g}(\mu)$
 
By construction,  all the solutions found in this way are finite at the origin. The solution divergent at the origin will be found by setting $\frac{1}{\widetilde {F}(0)}=0$ in eq. (\ref{SDRcst3}). It can also be approached by making $\widetilde {F}(0)$ larger and larger.

We have looked for solutions of eq.(\ref{SDRcst3}) with the integral cut in the UV at $q=30$~GeV. We have discretized it in $k$ and
 $q$. Taking values of the momenta spaced out by
$0.01$~GeV for $q\leq 2$~GeV and by $0.1$~GeV for $q\geq 2$~GeV we have computed the angular integral of the r.h.s. of
eq.(\ref{SDRcst2}). Then,  we solved this equation by iteration. Minus the integral in the r.h.s. is positive,  allowing an easy convergence. We linearize it at each step,   following the Newton method,  to accelerate the convergence of the iteration procedure,  as suggested by Bloch.
\vskip 0.3 cm
The results are the following:
\vskip 0.3 cm

1) \textbf{Critical case,  singular solution}. We find a solution with {{\penalty 10000} $\frac{1}{\widetilde {F}(0)}=0$},  i.e. $\widetilde {F}(0)=\infty$. We find then the corresponding "critical" constant:
\bea 
\widetilde {g_c}^2=\widetilde {F}(1.5~{\mathrm{ GeV}})= 33.198.... 
\eea
The relation of eq.(\ref{g2c}) happens to be very well satisfied:
\bea
\widetilde {g_{c}}^2  A^2 G^{(2)}(0) \frac{1}{10\pi^2}  \thickapprox 0.994....
\eea
\vskip 0.3 cm
The integration near $k=0$ can be improved by taking explicitly into account the analytical behavior of the kernel,  and assuming that the solution behaves as $1/k$ at small $k$. This imposes eq. (\ref{g2c}),  and one checks that 
$\widetilde {g}^2_c  k^2 F(k^2)^2 G^{(2)}(k^2)\frac{1}{10\pi^2}$ goes very smoothly to $1$
when $k \to 0$.
\vskip 0.3 cm

\begin{figure}[hbt]
\begin{center}
%\leavevmode
%\includegraphics[height=8cm]{/home/leyaouan/dir-Quadrics/dir-gluons/dir-cptIR/figSD.eps}
%\mbox{\epsfig{file=figSD.eps, height=8cm}}
\includegraphics[height=8cm]{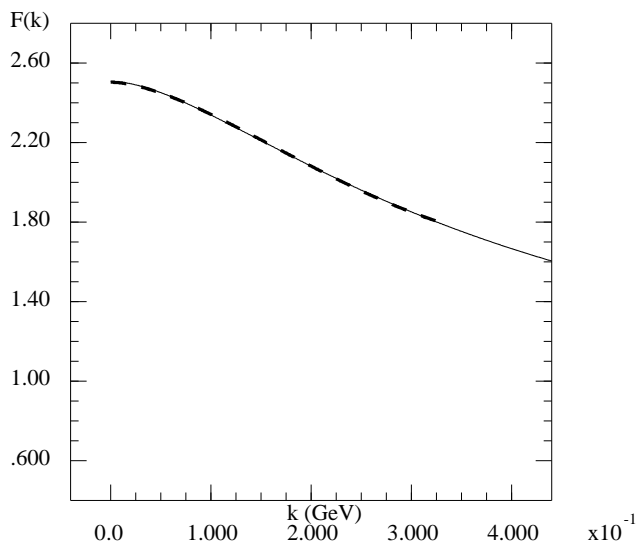}
\vskip -0.5 cm
\caption{\small The  $a+b k^2 \log(k^2)$ fit  at small momentum (dashed line) to our continuum SD prediction for the ghost dressing function, renormalised at $\mu=1.5$~GeV for $\widetilde g^2=29.$(solid line) ; the slope of the $k^2 \log(k^2)$ term is $4.06$ ; the agreement with the expected coefficient of $k^2 \log(k^2)$,  $4.11$ from the eq. (\ref{regular}),  is striking}
\label{pente}
\end{center}
\end{figure}

2) \textbf{Regular case}. We find a solution for all $\widetilde {F}(0)>0$,  and only one for each $\widetilde {F}(0)>0$ with our method of solution. . From our numerical solution at $\widetilde g^2 \simeq 29$,  which corresponds to the best description of lattice data (see Fig. \ref{fantomefig}), we can test the relation (\ref{regular}) giving the $k^2 \log(k^2)$
term. The result is presented in Fig.\ref{pente},  and the slope agrees well with what is expected from (\ref{regular}): $4.06$ against $4.11$.

The critical value of the coupling constant,  as well as the corresponding curve of $\widetilde {F}(k)$,  can be very well approximated by the regular
solutions at very large $\widetilde {F}(0)$. When $\widetilde {F}(0)$ is larger and larger, the eq. (\ref{regular}) is valid only in a smaller and smaller region near $k=0$, while in an intermediate region, we observe a $1/k$ behaviour. 

\vskip 0.5 cm

In conclusion,  in the case $\alpha_G=1$ we have exhibited a continuum set of IR finite solutions for arbitrary $F(0)$,  
and a unique singular solution for $\widetilde {g}^2=\widetilde {g}^2_c$,  with $\alpha_F=-\frac{1}{2}$. These
are the only solutions obtained with our iteration procedure using the Newton method. Of course this
doesn't prove that no other solution exists.

\subsection{Why the regular solutions have not been obtained previously} 

The regular solutions could not be obtained  by the proponents of the equation \ref{IRstable},  because,  as it seems to us,   they discard them 
from the beginning,  and thereby choose the critical value of the coupling constant,  by making an implicit assumption when solving  the so-called "infrared equation" for the ghost SD equation.
One can see this in the papers by R. Alkofer et al. (for instance \cite{alkofer-AnnPhys1998}, eqs. (43),  (44)),  or in the detailed discussion of Bloch \cite{bloch-ph-0303125}),  eqs. (55) to (58) .

Let us explain this briefly. They consider the above unsubtracted equation (note that this requires then an UV cutoff,  which we avoid in our discussions by considering the subtracted form--see below,  next section--) ; we write again the unsubtracted form :
\begin{eqnarray} 
\frac{1}{F_R(k^2)}=\widetilde Z_3-N_c g_{R}^2 \widetilde z_1 \int\frac{d^4 q}{(2\pi)^4}
\left(1- \frac{(k.q)^2}{k^2 q^2} \right)  \nonumber \\
\left[ \frac{G_R((q-k)^2)H_{1R}(q, k)}{((q-k)^2)^2}\right] F_R(q^2) \label{Rsubtracted}
\end{eqnarray}

One tries to match the small $k^2$ behaviour on both sides of eq.  \ref{Rsubtracted}. This is done for example in eq. (58) of \cite{bloch-ph-0303125}. A condition is then written which consists in equating the coefficient of $(k^2)^{-\alpha_F}$ with the corresponding one in the r.h.s.. However,  one notices that on the r.h.s..,  there is a constant contribution  $\varpropto(k^2)^0$. Therefore  unless the constant term  $\tilde Z_3$ is cancelled by the integral contribution for $k \to 0$,  we have necessarily $\alpha_F=0$. To have $\alpha_F<0$ as the author finds,  one needs this cancellation. This is what is \textbf{implicitly assumed},  but not stated explicitly. The condition of cancellation is :
 \begin{equation}
\widetilde Z_3=  \\
N_c g_{R}^2 \widetilde z_1 \int\frac{d^4 q}{(2\pi)^4}
\left.\left(1- \frac{(k.q)^2}{k^2\,q^2} \right)F_R(q^2)\right\arrowvert_{k=0}
\end{equation} 
However,  this additional equation {\bf does not derive} from the starting SD ghost equation,  and indeed it is not satisfied in general by the solutions of this basic equation,  as we show by displaying actually IR finite solutions. In fact,  it  can be valid only for a \textbf{particular value of the coupling constant,  the critical one} which is
 solution to the equation of Bloch,  his  eq. (58),  and which we derive rigorously through the subtracted equation (see our eq.\ref{g2c}). 

Let us make precise that at this stage the value of the coupling constant is taken as a free parameter.
It should be however fixed in the end consistently with the lattice data which we are using, which is done in the next section.

\section{Phenomenology} 

Having presented the general study and found the announced two types of solutions for the ghost dressing function,  either regular or singular,  the question is : which one is effectively realised on the lattice,  and therefore in true QCD ? Let us recall the classical problems which hamper the answer :
it is not possible  to know from the lattice data with total certitude whether the ghost dressing function is singular or not,  because 1) on the one hand,  the "singular" qualification in itself does not tell how close one should be to the zero momentum for the singularity to show up ; 2) on the other hand,  one cannot get arbitrarily close to zero momentum on the lattice.

\begin{figure}[hbt]
\begin{center}
%\leavevmode
%\includegraphics[height=8cm]{/home/leyaouan/dir-Quadrics/dir-gluons/dir-cptIR/figSD.eps}
%\mbox{\epsfig{file=figSD.eps, height=8cm}}
\includegraphics[height=8cm]{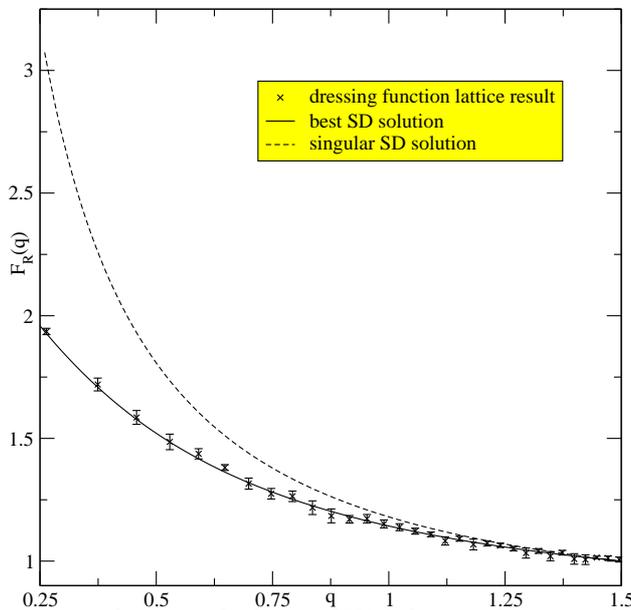}
\vskip -0.5 cm
\caption{\small Comparison between the lattice SU(3) data at $\beta=5.8$ and with  a volume $32^4$ for the ghost dressing function and our continuum SD prediction renormalised at $\mu=1.5$~GeV for $\widetilde g^2=29.$ (solid line) ; the agreement is striking ; also shown is the singular solution which exists only  at $\widetilde g^2=33.198....$ (broken line),and  which is obviously excluded.}
\label{fantomefig}
\end{center}
\end{figure}

 A better mean is offered by our calculation : it predicts the behavior of the respective solutions for the ghost over the \textbf{whole range} of momentum,  and not only very close to $k=0$ ; then looking to the lattice data,   we can  identify which is the more compatible with the data \footnote{ At this stage,  it is useful to stress the advantage of working with the renormalised form of the SD equations ; indeed the continuum and lattice versions are more directly comparable than the bare ones. As we have seen in our paper \cite{nous-ph-0507104v4},  the bare lattice equation for the ghost is affected by an important artefact which vanishes only very slowly with the cutoff,  being of order ${\cal O}(g^2)$. In the renormalised version,  this effect is included in the renormalisation constant $\widetilde Z_3$,  and we are left only with the much smaller cutoff effects of the type ${\cal O}(a^n)$}.
From fig. \ref{fantomefig},  we see that we can discard rather safely the singular solution,  because it is
 passing 
 much above the lattice data points over a {\bf very large range of momentum}: around $50 \%
 $ at the leftmost point measured on our lattice, $k=0.26$~GeV,  but still quite sizable near $k=0.5$~GeV. It is therefore quite unprobable that any lattice IR artefact could fill the gap. The advantage of our method is that,  by calculating what the critical solution  should be at rather large momenta,  we are able to discard it more convincingly.

On the other hand,  we find a very good description of the lattice data in the range $\widetilde g^2=28.3-29.8$ (the range is defined by one standard deviation except for the lowest point). This striking agreement is illustrated by Fig. \ref{fantomefig} (For indication, we quote the IR limit $F_R(0)=2.51$ for the same $\mu=1.5$ and $\widetilde g^2=29$). 
\vskip 0.5 cm

Moreover,  we can perform the following consistency test. We start from the equation (\ref{g2eff}) defining $\widetilde {g}^2$ in terms of bare quantities and we apply it to connect the continuum  $\widetilde {g}^2$ to the lattice bare quantities:
 \begin{equation} \label{g2eff2}
\widetilde {g}^2=N_c g_R^2 \widetilde z_1=N_c (6/\beta)~F_B^2(\mu^2) G_B(\mu^2)H_{1B} 
\end{equation}
We then ask whether our range $\widetilde g^2=28.3-29.8$ is reasonably consistent with the r.h.s. of eq.(\ref{g2eff2})as given by lattice data.
Let us stress that eq. (\ref{g2eff2}) should be then only approximate in several respects; first,  it is valid up to finite cutoff effects,  as well as volume effects ; second,  we have replaced the lattice vertex invariant $H_{1B}(q, k)$ by the constant
$H_{1B}$,  which is very rough ; and we cannot test on the lattice the assumed constancy over the momenta which are actually implied in our calculation. We have only at our disposal a lattice measure
of $H_{1B}$  at $q=k$. Last but not least,  lattice measurement of such vertex quantities is difficult; it is very noisy. Then, the test is only qualitative; nevertheless,  the result is very encouraging, as we see now.

Indeed, from the  above value of $\widetilde g^2$ found in the continuum on the one hand and, on the other hand, from the lattice data  $\beta=5.8$, $G_B(\mu^2)\simeq 2.89$ and $F_B(\mu^2)\simeq 1.64$ ($\mu$ is here chosen as $1.5$~GeV),  we can deduce the value of the factor $H_{1B}$ needed to satisfy equation (\ref{g2eff2}), and which represents some average on momenta. We find $H_{1B}\backsimeq 1.2$. This number should be compared  to the lattice measurements which are for $H_{1B}$  at $q=k$, and which give about $1.$,with large errors  (slightly larger or equal to $1.$, according to the $k$ value,  see ref.\cite{sternbeck-lat-0511053}). The comparison seems very encouraging considering the large uncertainties of the procedure : lattice artefacts, errors on $H_{1B}$, and finally the fact that our $H_{1B}$ is some average on momenta away from $q=k$.

\begin{figure}[hbt]
\begin{center}
%\leavevmode
%\includegraphics[height=8cm]{/home/leyaouan/dir-Quadrics/dir-gluons/dir-cptIR/figSD.eps}
%\mbox{\epsfig{file=figSD.eps, height=8cm}}
\includegraphics[height=8cm,angle=-90]{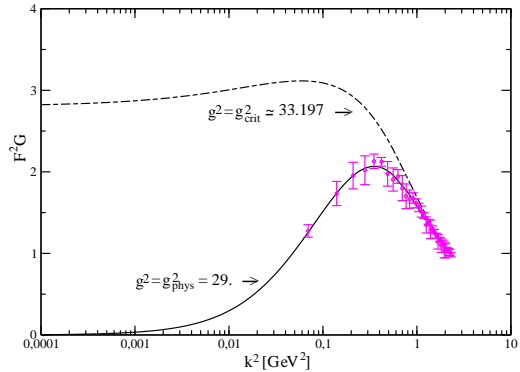}
\vskip -0.5 cm
\caption{\small Comparison between our lattice SU(3) data at $\beta=5.8$ and with  a volume $32^4$ for the product of gluon times ghost square dressing functions $G_R(k) F_R(k)^2$, renormalised at $\mu=1.5$~GeV, and the corresponding curve for the continuum singular solution $\alpha_G+2 \alpha_F=0$, which exists only  at $\widetilde g^2=33.198....$;
it is obviously excluded; up to a factor $g^2/4 \pi$, it corresponds to the $\alpha(k^2)$ presented in Fig.8 by Fischer \cite{fischer-ph-0605173v2}; the form is very similar. Also shown is our continuum regular solution for $\widetilde g^2=29.$ (solid line) for which the agreement is striking.}
\label{GF2fig}
\end{center}
\end{figure}

Another way of presenting the striking difference between the regular solution and the singular one is in term of the familiar product discussed in the introduction : $G_R(k) F_R(k)^2$ \footnote{Usually, this quantity is presented with multiplication by an additional factor including the renormalised coupling constant and possibly other factors. Here, we present the raw product to avoid any ambiguity in such procedures.}. From
 the analytical discussion, in the {\bf critical} case, it should tend to $10~\pi^2/\widetilde{g_c}^2 \simeq 3.$ when $k \to 0$ , and numerically it should be $3.14$ at our smallest lattice momentum $k=0.26$ . This is completely at odds with the lattice : the lattice value is $1.28$ at $k=0.26$, with a clear tendency to still lower values at smaller $k$. On the contrary, our regular solution fits perfectly the lattice data. We illustrate this in Fig. \ref{GF2fig}. Let us stress that our SD solutions (continuous curve) are obtained {\bf in the continuum} and {\bf in infinite volume} ; they appeal to the lattice data only to have a physically reasonable definite gluon propagator as input to the SD equation. 

\vskip 0.5 cm

 \section{Conclusion} 

The relation $\alpha_G+2 \alpha_F=0$ is usually
believed to be an unavoidable consequence of SD equations. The  problem is that the lattice data grossly contradict this. Indeed, 
for small momenta $G(k^2) F(k^2)^2 \to 0$ very fast. We resolve this contradiction in the following way.   We show that this belief is wrong and that an alternative exists by solving numerically the ghost SD equation with input from the lattice for the gluon propagator and the vertex . 

The alternative is the one we have previously envisaged as a possibility in a general analytical analysis \cite{nous-ph-0507104v4, nous-ph-0604056}: $\alpha_F=0$ with $\alpha_G \geqslant 1$,  therefore $G(k^2) F(k^2)^2 \to 0$ as shown by the data,  but in the present article,  the existence of such a solution is demonstrated by actually solving in $F$ the equation for a given $G$.   This solution violates the statement $\alpha_G+2 \alpha_F=0$.

The relation $\alpha_G+2 \alpha_F=0$ would imply that the ghost factor is singular,  since $\alpha_G>0$. 
The numerical solution of the equation then adds another strong reason for rejecting this relation. A singular solution -- which necessarily satisfies  $\alpha_G+2 \alpha_F=0$ -- only exists for a definite value of the coupling constant. We calculate it and find that it grossly differs
from the lattice data on a large range of $k$,  and not only for the smallest momenta.

The alternative solution which is regular ($\alpha_F=0$), is realised if the coupling constant is smaller than a certain critical value,  while the singular one is present only at the critical value.  The lattice data for the ghost are very well reproduced for a coupling constant close to the one expected from the value
 $\beta=6.0$ used for the lattice calculation. We are then confident to have found the actual QCD solution,  up to moderate artefacts.

Our numerical study therefore adds strong new arguments from the lattice data in favor of this alternative . It is based on the ghost equation only,  since we feel that the gluon equation,  being  much more complicated,  suffers from much more uncertainties,  due to the necessary critical approximations to be made - this has been illustrated by the findings of Bloch \cite{bloch-ph-0303125}. 

The alternative, regular, solution has not been found in usual studies,  because they have chosen by construction the critical value.

The important physical consequence is that we do not get the alleged non trivial IR fixed point for the MOM coupling constant since $g_R(k) \to 0$ when $k \to 0$ at fixed $g_0$ as for the three-gluon couplings,  in agreement with lattice data. At this point,  it is important to insist on the fact that there are infinitely many  definitions of "the" QCD coupling constant. A priori,  there is no reason for \textbf{their IR behaviour to be universal}. 
In particular, the  ones defined from elementary fields Green functions have no reason  to behave in the same way as more physical definitions such as taken from the perturbative expansion of certain physical hadronic amplitudes.

\section*{Acknowledgements} We would like to thank Alexei Lokhov, who gave the impulse to our research on the SD ghost equation.

\end{document}